\documentclass[finalnew]{agujournal2019}
\usepackage{url} 
\usepackage{lineno}
\usepackage[finalnew]{trackchanges} 
\usepackage{soul}
\usepackage{natbib}
\usepackage[none]{hyphenat}
\usepackage{amsmath}
\usepackage{amssymb}

\usepackage[iau]{journalnames}
\draftfalse

\journalname{JGR: Planets}

\begin{document}

%
%

\title{1I/`Oumuamua as an N$_2$ ice fragment of an exo-Pluto surface: I. Size and Compositional Constraints}

%
%




\authors{Alan P. Jackson\affil{1}\thanks{ORC-ID 0000-0003-4393-9520}, Steven J. Desch\affil{1}\thanks{ORC-ID 0000-0002-1571-0836}}


\affiliation{1}{School of Earth and Space Exploration, Arizona State University, Tempe, AZ 85087}




\correspondingauthor{Alan P. Jackson}{alan.jackson@asu.edu}




\begin{keypoints}
\item A fragment of N$_2$ ice would satisfy all known constraints on `Oumuamua's size, composition, and acceleration.
\item The presence of N$_2$ ice in the solar system on Pluto makes it likely `Oumuamua is such a fragment
\item The existence of `exo-Plutos' in other stellar systems must be common.
\end{keypoints}

%
%

%
%

\justify
\begin{abstract}
The origin of the interstellar object 1I/`Oumuamua has defied explanation. We perform calculations of the non-gravitational acceleration that would be experienced by bodies composed of a range of different ices and demonstrate that a body composed of N$_2$ ice would satisfy the available constraints on the non-gravitational acceleration, size and albedo, and lack of detectable emission of CO or CO$_2$ or dust.  We find that `Oumuamua was small, with dimensions 45~m $\times$ 44~m $\times$ 7.5~m at the time of observation at 1.42~au from the Sun, with a high albedo of 0.64.  This albedo is consistent with the N$_2$ surfaces of bodies like Pluto and Triton. We estimate `Oumuamua was ejected about 0.4-0.5 Gyr ago from a young stellar system, possibly in the Perseus arm. Objects like `Oumuamua may directly probe the surface compositions of a hitherto-unobserved type of exoplanet: ``exo-plutos''. In a companion paper (Desch \& Jackson, 2021) we demonstrate that dynamical instabilities like the one experienced by the Kuiper belt, in other stellar systems, plausibly could generate and eject large numbers of N$_2$ ice fragments.  `Oumuamua may be the first sample of an exoplanet brought to us.
\end{abstract}

\section*{Plain Language Summary}
1I/`Oumuamua is very strange and it is hard to explain where it came from.  We looked at several different ices and the push they would give `Oumuamua as they evaporated.  We found that the best ice is nitrogen (N$_2$), which would explain many of the things we know about it.  `Oumuamua was small, about half as long as a city block and only as thick as a three story building, but it was very shiny.  The shininess is about the same as the surfaces of Pluto and Triton, which are also covered in nitrogen ice.  We suggest `Oumuamua was probably thrown out of a young star system about half a billion years ago.  Bodies like `Oumuamua may allow us to see what the surfaces of a so far unknown type of exoplanet, ``exo-Plutos'', are made of.  In a following paper (Desch \& Jackson, 2021) we show that orbital instabilities in which giant planets move around, as happened in our own outer solar system 4 billion years ago, could make and throw out large numbers of small pieces of nitrogen ice like `Oumuamua.  `Oumuamua may be the first piece of an exoplanet brought to us.

%
%

\section{Introduction}
\label{sec:intro}

1I/`Oumuamua was discovered by the Pan-STARRS telescope in October 2017, when it was only 0.22~au from Earth and briefly “brightened” to a 20th-magnitude object. Its heliocentric orbit was soon found to be hyperbolic, with eccentricity $e=1.2$, making `Oumuamua the first definitive interstellar object discovered \citep{meech2017}. In August 2019, a second object on a hyperbolic orbit ($e=3.36$) was discovered from a home observatory, the interstellar comet 2I/Borisov . These two objects have been travelling through our solar system for centuries and must be part of a population of millions of such interstellar objects currently passing through the Solar System. This newly discovered population of objects provides an opportunity to probe the compositions and physical properties of analogs to comets and asteroids in extrasolar systems. In addition, they offer the opportunity to test whether the same processes that occurred in our solar system have occurred in other planetary systems. 

These two interstellar objects appear to very different.  2I/Borisov is readily recognised as a comet like those in our own solar system, actively outgassing species like CN \citep{fitzsimmons2019}.  1I/`Oumuamua on the other hand appears very different from solar system comets and has been much more resistant to explanation. Several oddities of `Oumuamua were compiled by \citet{bialy2018}, although most of its unusual properties are not as mysterious as they seem at first glance, as made clear in the review by \citet{oumuamuateam2019}. We review these here.

One of the most striking features of `Oumuamua is its extreme axis ratio. Initial calculations based on the observed light curve suggested a prolate spheroid with a length as much as 10 times its width \citep[e.g.,][]{meech2017}.  Subsequent, more detailed, examination of the spin state by \citet{belton2018} and fits to the lightcurve data by \citet{mashchenko2019} found that both prolate and oblate solutions are possible, with axis ratios in the range of 5:1 to 10:1, preferring the oblate solution.  Even the low end of this range is substantially higher than the axis ratio of any known solar system object and thus appears extremely unusual.  Such an extreme axis ratio is, however, consistent with erosion over time, and \citet{domokos2009} have shown that erosion of the surface of small bodies can increase the aspect ratio of objects into this range.

The upper limits on the thermal emission from `Oumuamua (from \emph{Spitzer Space Telescope} observations) indicate a body no more than a few hundred meters in size and suggest an albedo higher than asteroids or comets in the solar system \citep{trilling2018}. This non-detection by \emph{Spitzer} sets an upper limit on the diameter (if spherical) of 98-440~m, depending on model assumptions \citep{meech2017, jewitt2017}, which translates into $V$-band albedos $>$0.2 to 0.01 for the same size ranges, indicating a value that is likely larger than typical for comets and asteroids ($<$0.1), although hardly unphysical.

It has been inferred -- from calculations of the Pan-STARRS detection probability and the length of time the observatory has been operational -- that the detection of `Oumuamua implies that each star in the galaxy must have ejected $\sim 10^{15} - 10^{16}$ such objects \citep{do2018}, several orders of magnitude higher than predictions made prior to the discovery of `Oumuamua \citep[e.g][]{jewitt2003, moromartin2009}.  Our knowledge of the size-frequency distribution of objects less than 1~km in size has always been limited however, and $10^{15} - 10^{16}$ objects ejected per solar system is not implausible.  In addition our knowledge of the occurrence rate of exoplanets was dramatically improved since those earlier predictions of the numbers of interstellar objects.  At any rate the true number density is only constrained to a wide range, $3.5\times10^{13} - 2\times10^{15}$~pc$^{-3}$ \citep{portegieszwart2018}, or $3\times10^{14} - 2\times10^{16}$ per M$_{\odot}$ of star.

The velocity of `Oumuamua with respect to the local standard of rest (LSR), the average velocity of stars in the neighbourhood of the Sun, was only 9~km~s$^{-1}$ \citep{meech2017}, far less than the tens of km~s$^{-1}$ average dispersion of stars with respect to the LSR, and unexpected if `Oumuamua were ejected from an average stellar system.  While such a low velocity with respect to the LSR is not common among all stars, it is common among relatively young ($<$2~Gyr-old) stellar systems.  Stars are born from molecular clouds, which have a typical velocity dispersion of $\sim$6~km~s$^{-1}$ with respect to the local standard of rest, and acquire greater velocity dispersions over time through stellar encounters.

Most of the mysteries of `Oumuamua appear to have prosaic explanations, but the most enduring mystery about `Oumuamua regards its composition and, related to it, its non-gravitational acceleration.  Observations placed strict upper limits on the outgassing rates of dust, CO and CO$_2$ \citep{jewitt2017, trilling2018}.  With no direct observations of outgassing, initial work focused on the idea of a very volatile-poor body, and explaining how such a body might originate \citep[e.g.][]{cuk2018,jackson2018}.  Further observations, however, revealed that the trajectory of `Oumuamua could not be fully explained by an object moving purely under the action of the Sun's gravity.  Instead, explaining the motion of `Oumuamua required an additional non-gravitational force directed outward, at a level about 10$^{-3}$ that of the gravitational force, and varying as roughly $1/d^2$ (where $d$ is distance from the Sun) \citep{micheli2018}.  This would be consistent with cometary outgassing, but appears at odds with the strict upper limits on species that are typically found in cometary comae.  It has not been clear what ice composition could provide sufficient force through sublimation to explain the non-gravitational acceleration, while simultaneously remaining undetectable.

Any analysis of the force due to sublimation of ices starts with an estimate of `Oumuamua’s size and shape. The light-curve of `Oumuamua while it was around 1.4~au from the Sun was carefully analysed by \citet{mashchenko2019}, including the possibility of torques induced by non-isotropic ejection of material (as might be expected to accompany the observed non-gravitational acceleration).  This analysis produced two possible best-fit ellipsoids: one prolate (cigar-shaped), with axis ratios of roughly 8:1:1; and one oblate (pancake-shaped), with axis ratios of roughly 6:6:1.  The oblate solution is favoured on statistical grounds because the prolate solution requires fine-tuning of the orientation of the angular momentum vector.  An oblate shape was also favoured by \citet{sekanina2019} on the basis of the {\it Spitzer} non-detection.  \citet{seligman2020} combined the shape derived by \citet{mashchenko2019}, the average illumination experienced by `Oumuamua, and assumptions about its composition, to determine the magnitude of the non-gravitational acceleration. A prolate ellipsoid cannot experience the required acceleration unless it has substantial amounts of H$_2$ ice, but an oblate spheroid shape is more accommodating of other compositions, further favouring the oblate scenario. We consider the shape favoured by \citet{mashchenko2019} and \citet{seligman2020}: axes (diameters) $a$=115~m, $b$=111~m, $c$=19~m (assuming an albedo 0.1), and assumed average projected-area-to-surface-area ratio $\xi \approx$0.19 (during the times of the light curve observations).

\citet{seligman2020} proposed that `Oumuamua contained a substantial quantity of H$_2$ ice, which they showed could provide the necessary non-gravitational acceleration if it covered around 6\% of the surface.  This is problematic, however, because H$_2$ has a condensation temperature of $<$10~K that is only reached in the most extreme molecular cloud core environments \citep{kong2016}.  While more volatile substances can be entrained in common ices like H$_2$O and CO$_2$, the release of these entrained volatiles is largely controlled by the sublimation of the less volatile parent ice \citep[e.g.][]{fayolle2016}.  In this work we focus on a possibility that was overlooked by \citet{seligman2020}: N$_2$ ice.  They found that a surface composed primarily of N$_2$ ice would provide sufficient acceleration and, in contrast to hypothetical H$_2$ ice, significantly pure N$_2$ ice \emph{is} observed in the solar system, on the surfaces of Pluto and Triton.

The surface of Pluto is $> $98\% N$_2$ ice, with frosts of CH$_4$ and CO comprising the remainder \citep{protopapa2017}. The N$_2$ ice today is concentrated in and fills the Sputnik Planitia impact basin, which is at least 2-3~km, possibly up to 10~km, deep. The N$_2$ ice flows into the basin as glaciers, and then convects to transport the heat flux rising up from below it; from the lateral extent of the convection cells it is inferred that the N$_2$ ice is at least several km thick \citep{mckinnon2016}. The amount of ice in Sputnik Planitia is equivalent to a global layer 200-300~m thick \citep{mckinnon2016}, but could have been much larger in the past. Similarly, Triton’s surface today is dominated by global layer of N$_2$ ice estimated to be about 1-2~km thick \citep{cruikshank1998}.

In this paper we examine the hypothesis that `Oumuamua is a body composed of pure N$_2$ ice.  In Section~2 we consider the combinations of albedo and sizes that would be consistent with a pure N$_2$ ice body, and calculate its size and mass before entering the solar system.  In Section~3 we consider the constraints on `Oumuamua's size and shape and its survival through the interstellar medium.  In Section~4 we compare how well this hypothesis compares with others, such as `Oumuamua being a more typical comet, or a body rich in H$_2$ ice.

\section{The non-gravitational acceleration for different compositions}
\label{sec:nongrav}

Alongside H$_2$, \citet{seligman2020} presented calculations of the non-gravitational acceleration for a number of other ices, including N$_2$, which we focus on here.  However, we wish to revisit some of the assumptions that they made in their calculations and present a revised formulation of the non-gravitational acceleration.  Their Equation 1 for the flux of sublimated molecules is highly simplified and does not include effects like the influence of evaporative cooling on surface temperature or the jetting velocity in the fluid regime.  In addition, they assume throughout that the albedo of `Oumuamua is 0.1, which is not necessarily appropriate for the ices under consideration.

\subsection{Size and albedo}
\label{sec:nongrav:size}

\citet{mashchenko2019} provided two possible solutions for the shape of `Oumuamua: their preferred oblate spheroid solution, with axes 115~m $\times$ 111~m $\times$ 19~m; and a prolate spheroid, with axes 342~m $\times$ 42~m $\times$ 42~m.  Both of these solutions, however, assume that `Oumuamua has a geometric albedo, $p_{\rm G}$, of 0.1.  Since the light reflected is proportional to $p_{\rm G}$, it follows that the axis lengths are proportional to $p_{\rm G}^{-1/2}$, since the shape is fit to a light curve of known brightness.  The mass of `Oumuamua (for a constant density) is thus proportional to $p_{\rm G}^{-3/2}$ and the mass-to-surface area ratio, denoted as $\eta$ by \citet{seligman2020}, is proportional to $p_{\rm G}^{-1/2}$.

The calculation by \citet{seligman2020} of the non-gravitational acceleration assumes these axes (and thus $p_{\rm G}$=0.1) and that the Bond albedo, $p_{\rm B}$, is also 0.1.  The radiation absorbed by `Oumuamua is proportional to $(1-p_{\rm B})$, as reflected in Equation 1 of \citet{seligman2020}.  Setting $p_{\rm G} = p_{\rm B} = p$, we can see that in Equation 6 of \citet{seligman2020} this leads to the non-gravitational force being roughly proportional to $(\sqrt{p}(1-p))^{-1}$, a quadratic with a maximum at $p = 1/3$.

It is clear that the choice of albedo is important for calculating the acceleration, but it is not clear what is the correct choice of albedo for exotic ices.  As such we choose to treat the albedo as a variable in our calculations, later comparing the values that satisfy the required non-gravitational force to known ice albedos.  As such, we also treat the size of `Oumuamua as a variable.  We keep the axis ratios of the \citet{mashchenko2019} oblate solution ($\sim$6:6:1), but allow the axis lengths to scale as $p_{\rm G}^{-1/2}$.  Note that the oblate solution of \citet{mashchenko2019} is very close to a perfect oblate spheroid and so \citet{seligman2020} approximated it as 113~m $\times$ 113~m $\times$ 19~m; we retain the full triaxial treatment, but the difference is small.

\subsection{Temperature}
\label{sec:nongrav:temp}

Whatever ice `Oumuamua is composed of, it is likely that around perihelion the body will be losing mass at very high rates, as can be seen from the necessary surface covering fractions computed by \citet{seligman2020}.  Material carries energy away from the body as it sublimates, so an accurate calculation of the surface temperature and the mass loss rates must include the contribution of evaporative cooling.  Following the treatment of \citet{hoang2020} for H$_2$ ice, we consider the energy balance between absorbed sunlight and the combination of thermal emission and sublimation:
\begin{equation}
\label{eq:energybalance}
    \frac{L_{\odot}}{4\pi d^2}\, \xi_0 \, (1-p_{\rm B}) = 
    \epsilon \, \sigma T_{\rm surf}^4
    - n \left[\Delta H_{\rm sub} + \Delta H_{\rm trans} + m C_p \Delta T\right] \frac{dR}{dt}.
\end{equation}
Here $\xi_0$ is the ratio of the area projected to the Sun to the total surface area.  While $\xi_0$ was 0.19 over the (relatively short) epoch of the light curve observations \citep{seligman2020}, we assume that the isotropic value of 0.25 applied over longer periods, especially given the tumbling rotation of `Oumuamua \citep{fraser2018}.  The infrared emissivity of the ice is $\epsilon$, which we assume is roughly 0.85, an appropriate value for N$_2$ ice \citep{stansberry1996}.  The number density of molecules (e.g., N$_2$) in the ice and the mass of a molecule are $n$ and $m$, respectively, and related by $n = \rho/m$, where $\rho$ is the mass density of the ice.  The temperature difference between the interior and the surface is $\Delta T = T_{\rm surf}-T_{\rm int}$, while $C_p$ is the heat capacity of the ice.  We consider two enthalpies: $\Delta H_{\rm sub}$ is the enthalpy of sublimation, while $\Delta H_{\rm trans}$ is the enthalpy associated with any solid-state phase transitions that occur between $T_{\rm int}$ and $T_{\rm surf}$.  Finally, $dR/dt$ is the rate of change in the radius of `Oumuamua, which is also temperature-dependent and given by
\begin{equation}
\label{eq:drdt}
    \frac{dR}{dt} = - \frac{\nu}{n^{1/3}} \exp\left(\frac{-\Delta H_{\rm sub}}{kT_{\rm surf}}\right),
\end{equation}
where $\nu$ is the lattice vibrational frequency and we assume $dR/dt$ is the same in all directions so that mass is lost isotropically.

We highlight two necessary simplifying assumptions that we have used.  The first is that mass is lost isotropically.  In reality there will be impurities in the material and irregularities in the structure that will make some parts of the body more susceptible to mass loss than others.  In addition, even assuming a perfectly regular body with no impurities, mass loss will be a function of the solar intensity at the relevant location on the surface and how long that location has been exposed to sunlight, and as such it will vary strongly over the surface of the body.  We would expect that the tumbling rotation of `Oumuamua should cause any variations in the amount of mass lost from different parts of the body to average out, which leads into our second simplifying assumption.  We have assumed that the appropriate value of $\xi_0$ is the isotropic value of 0.25 and over the whole of `Oumuamua's journey through the solar system the tumbling rotation should indeed average the projected area out to the isotropic value, but as we can see from the epoch of the light curve observations $\xi_0$ can deviate significantly from this expected value in shorter time intervals; if such deviations occur during periods of high mass loss, this could change the total mass lost during passage through the solar system and cause some parts of the surface to lose more mass than others.  Since the rotation is chaotic, however, it is not possible to model these presumably slight effects, and the tumbling probably justifies the assumption of isotropy.  

The final term in the square brackets in Equation~\ref{eq:energybalance} depends on the temperature difference between the surface and the interior and thus we need to determine what the interior temperature of the body will be.  During the very long period that `Oumuamua spent in interstellar space we can assume that the interior of the body reached the same temperature as the surface and that that temperature was in equilibrium with energy absorbed from the cosmic microwave background and cosmic rays such that
\begin{equation}
\label{eq:Tint}
    \epsilon \, \sigma T_{\rm CMB}^4 + F_{\rm GCR} = \epsilon \, \sigma T_{\rm int}^4.
\end{equation}
Assuming that $T_{\rm CMB}=2.73$~K, $\epsilon = 0.85$, and $F_{\rm GCR}=1.9\times10^{-2}$~erg~cm$^{-2}$~s$^{-1}$ (see Section~\ref{sec:temporal:ism} for details) we arrive at an interior temperature prior to encountering the solar system of 4.6~K.  Astrophysical ices have generally low thermal diffusivities. For example, N$_2$ ice has a thermal diffusivity of $\kappa=k/(\rho C_p) \sim 2.4\times10^{-7}$~m$^2$~s$^{-1}$, and with the exception of H$_2$ all of the other ices are within roughly a factor of 10 of this (see Table A1).

Considering that `Oumuamua was only inside a few au for a time $t \sim$ few $\times 10^7$~s before perihelion, the surface temperatures would have penetrated only to depths $\sim \sqrt{\kappa t} \sim 2$~m, similar to conclusions reached by \citet{fitzsimmons2018} and \citet{seligman2018}.  We can also cast this in a different way and ask how long it would take a heat pulse to penetrate to a depth of around 10~m, which as we will see later is approximately the size we will calculate for the shortest axis of `Oumuamua at perihelion.  For N$_2$ ice it would take about 6 years for a heat pulse to propagate to depths of around 10~m.  About 6 years before it was observed, `Oumuamua was at about 45~au from the Sun, at which distance sublimation would have been negligible and the surface temperature would have been around 25~K (from the balance of incoming and outgoing radiation).  At larger heliocentric distances `Oumuamua was likely moving slowly enough for surface heat to diffuse into the interior and make it approximately isothermal.  As such we make the assumption that the interior temperature of `Oumuamua was $T_{\rm int} = {\rm min}[25 \, {\rm K}, T_{\rm surf}]$.

A more accurate calculation would of course consider the propagation of heat from the surface to the interior in parallel with sublimation loses from the surface, but the effects of variations in the $\Delta T$ term would not substantially alter the calculations of mass loss, since this term is at most about 10\% the size of the $\Delta H_{\rm sub}$ term.

\subsection{Mass loss and non-gravitational acceleration}
\label{sec:nongrav:massloss}

The surface temperature of `Oumuamua and the rate of decrease in the radius are determined by iterating equations \ref{eq:energybalance} and \ref{eq:drdt}.  The mass loss rate is then easily obtained as $dM/dt = \rho \, S \, dR/dt$, where $S$ is the surface area of the body.  We then convert this into a force directed away from the Sun by assuming
\begin{equation}
    {\rm Force} = \frac{1}{3} \left(-\frac{dM}{dt}\right) V_{\rm jet},
\end{equation}
where $V_{\rm jet}$ is the effective jet speed of gas leaving the surface of the body and the coefficient 1/3 is a geometric factor accounting for the fact that the sublimation rate scales with the solar elevation angle, $\theta$, as $\cos \theta$, and the sunward component of the momentum of the sublimating gas is also reduced by a factor of $\cos \theta$.  Averaging over the sun-facing hemisphere yields a factor of 1/3 for a spherical body, and we assume that the same time-averaged value applies to a tumbling ellipsoid.

The form of the effective jet speed of the gas leaving the surface depends on whether we are in the free molecular flow or fluid regime.  Taking the example of N$_2$ we find that even at temperatures as low as 39~K the vapour pressure of N$_2$ exceeds 40~$\mu$bar \citep{frels1974}, implying a number density of N$_2$ molecules $> 10^{22}$~m$^{-3}$.  Assuming a molecular collision cross-section $\sim10^{-19}$~m$^2$ the mean free path of N$_2$ molecules does not exceed around 1~mm, much smaller than the size of `Oumuamua itself.  Other gases will produce similar values and so we assume that we are in the fluid regime.  As such we adopt the treatments of \citet{crifo1987} and \citet{maquet2012} for cometary outflows.  Specifically, we assume $V_{\rm jet} = \tau \sqrt{8k_B T_{\rm surf}/\pi m}$, where $\tau$ represents an averaging over velocity and depends on the Mach number of the outflow, but for typical Mach numbers observed in cometary outflows $\tau \approx 0.39-0.50$, and \citet{crifo1987} recommended $\tau \approx 0.45$.  For N$_2$ ice at a typical surface temperature of 25 K, this yields $V_{\rm jet} = 80 \, {\rm m} \, {\rm s}^{-1}$. This is similar to, but slightly different from, the jetting speed considered likely by \citet{seligman2020}, who used $V_{\rm jet} = \sqrt{\gamma k_B T/m}$, fixed the temperature at 25 K and $\gamma$ at $4/3$ (5/3 would be more appropriate), and found $V_{\rm jet}=99 \, {\rm m} \, {\rm s}^{-1}$ for N$_2$.

Rearranging Equation~\ref{eq:energybalance} we then obtain the mass loss rate,
\begin{equation}
\label{eq:dmdt}
    -\frac{dM}{dt} = S  m  \left[ \frac{L_{\odot}}{4\pi d^2}  \xi_0  (1-p_{\rm B}) - \epsilon  \sigma T_{\rm surf}^4 \right]  
    \left[ \Delta H_{\rm sub} + \Delta H_{\rm trans} + m C_p \Delta T \right]^{-1},
\end{equation}
and the non-gravitational acceleration, 
\begin{equation}
\label{eq:nongravacc}
    a = \frac{\tau}{3} \frac{S}{M} \sqrt{\frac{8 k_B m T_{\rm surf}}{\pi}} \left[ \frac{L_{\odot}}{4\pi d^2}  \xi_0  (1-p_{\rm B}) - \epsilon  \sigma T_{\rm surf}^4 \right] 
    \left[\Delta H_{\rm sub} + \Delta H_{\rm trans} + m C_p \Delta T \right]^{-1}.
\end{equation}

\subsection{Composition of `Oumuamua}
\label{sec:nongrav:comp}

\begin{figure}
    \centering
    \includegraphics[width=\columnwidth]{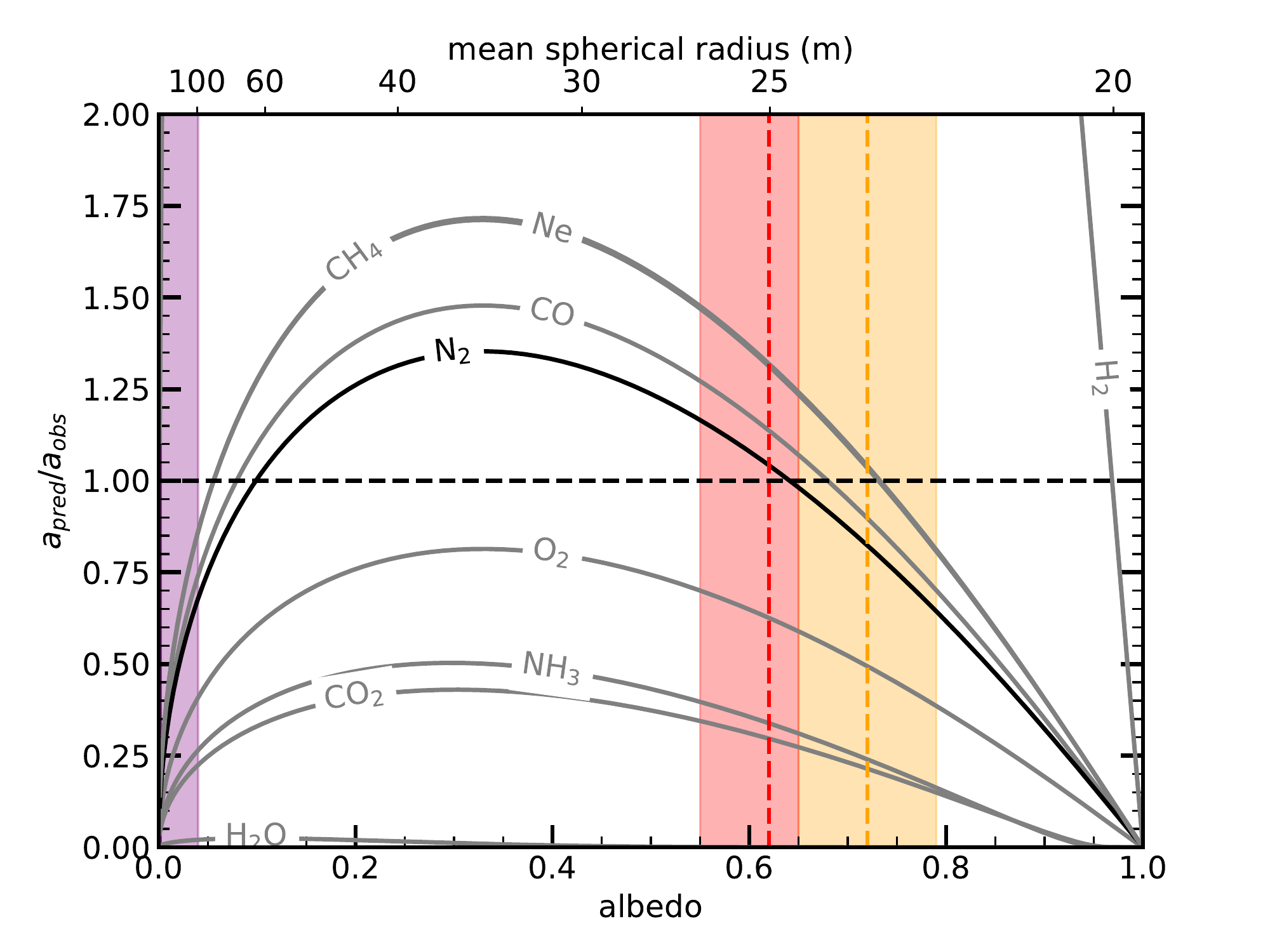}
    \caption{Predicted non-gravitational acceleration at 1.42~au due to sublimation and jetting, relative to the observed value, assuming `Oumuamua is an oblate ellipsoid of pure ice with the labelled compositions, for various values of the common geometric and Bond albedo.  The top axis converts albedo into equivalent mean spherical radius assuming a 6:6:1 axis ratio.  Note that the H$_2$ curve extends far above the plotted range, peaking at $\sim$13. The orange band shows the reported Bond albedo for Pluto ($p_B \approx 0.72 \pm 0.07$; \citet{buratti2017}), while the red band shows the R-band geometric albedo of Pluto reported by \citet{buratti2015}, including additional downward spread to account for uncertainty in whether the detect opposition surge should apply to ‘Oumuamua. The purple band shows the range disallowed by the size constraints from the Spitzer non-detection. The acceleration we predict for a chunk of N$_2$ ice matches `Oumuamua’s observed non-gravitational acceleration if its albedo is 0.64, consistent with the albedo of Pluto’s N$_2$ ice surface.}
    \label{fig:nongravalbedo}
\end{figure}

Having set out the equations that govern the mass loss and non-gravitational acceleration we are now in a position to compute the non-gravitational acceleration for a selection of different ices and compare the results to observations.  \citet{micheli2018} fit the non-gravitational acceleration as $4.92\times10^{-4} \, (d / 1 \, {\rm au})^{-2} \, {\rm cm} \, {\rm s}^{-2}$ over the observational arc that runs from \change[]{18}{14} October 2017 to 2 January 2018.  In Figure~\ref{fig:nongravalbedo} we plot the non-gravitational acceleration that we predict at 1.42~au, relative to the observed acceleration, as a function of albedo (assuming $p = p_{\rm G} = p_{\rm B}$) for a variety of pure ice compositions.  From the set of ices used by \citet{seligman2020} in their Table 1 we exclude Ar, Kr and Xe, as these heavy noble gases have low cosmic abundances and it seems highly unlikely that there would be large populations of bodies composed of these ices.  We add, however, the common astrophysical compounds CH$_4$, CO and NH$_3$, so that the complete set of ices we examine is H$_2$, Ne, CH$_4$, CO, N$_2$, NH$_3$, O$_2$, CO$_2$, and H$_2$O.  The sublimation enthalpies and other data for all 9 ices, along with the data sources, can be found in Table A1.

The abundant ices H$_2$O and CO$_2$ are immediately ruled out as the main constituents of `Oumuamua, as they are incapable of providing the necessary non-gravitational acceleration for any albedo, as was also found by \citet{seligman2020}.  Pure NH$_3$ and O$_2$ ices are less likely than H$_2$O and CO$_2$, and also ruled out as not capable of providing sufficient force.  The force from sublimation of Ne and H$_2$ would be sufficient to match the observed acceleration; indeed, H$_2$ provides so much acceleration that the curve does not easily appear in the same vertical scale as the other ices, and would in fact require `Oumuamua to have an albedo very close to 1 (an albedo very close to zero is ruled out by the \emph{Spitzer} observations).  Large accumulations of Ne and H$_2$ ices are unlikely to exist, however\change[]{,}{.}  \citet{hoang2020} enumerate many issues with forming an H$_2$ ice body, and with such a body surviving its journey from its origin to the Solar system.  With a sublimation temperature of only 9~K most of the issues that apply to H$_2$ also apply to Ne, along with the added problem of much lower cosmic abundance.  This leaves CH$_4$, CO and N$_2$ as the viable options.  CO is observed in comets and on Pluto, but any significant CO is ruled out by observations with \emph{Spitzer} \citep{trilling2018}.  CH$_4$ ices are observed on Pluto, including as dunes \citep{telfer2018}, but overwhelmingly CH$_4$ is observed as a trace species dissolved in N$_2$ ice, and never more than a few weight percent \citep{trafton2015}.  As such we identify N$_2$ ice as by far the most likely candidate for `Oumuamua's composition.  Figure~\ref{fig:nongravalbedo} shows that N$_2$ matches the observed non-gravitational acceleration for two values of the albedo: just above 0.1, and at 0.64.  The high-albedo solution is intriguing as this is close to the observed geometric and bond albedos of Pluto and Triton, both bodies that have large amounts of N$_2$ on their surfaces. For Pluto, \citet{buratti2015} find $p_{\rm G} = 0.62 \pm 0.03$ in R-band (red band in Figure~\ref{fig:nongravalbedo}), and \citet{buratti2017} find $p_{\rm B} = 0.72 \pm 0.07$ (orange band in Figure~\ref{fig:nongravalbedo}). For Triton, the equivalent values are $p_{\rm G} = 0.72-0.82$ in R-band \citep{hicks2004}, and $p_{\rm B} = 0.82 \pm 0.05$ \citep{hillier1990,nelson1990}.

Adopting geometric and bond albedos of 0.64, we infer that the dimensions of `Oumuamua were 45.5~m $\times$ 43.9~m $\times$ 7.5~m during the light-curve observations when it was located at 1.42~au from the Sun.  At that time its mass would have been $8.0 \times 10^6$~kg.

By the time of the \emph{Spitzer} observations on 21-22 November 2017 that place limits on the size and composition of `Oumuamua, it had receded to 2~au and would have shrunk slightly to 44.8~m $\times$ 43.2~m $\times$ 6.8~m.  The non-detection of `Oumuamua in thermal emission implies a diameter (if spherical) of $< 98-440$~m, a criterion clearly met by the size that we determine.  At this time we calculate that `Oumuamua would have been losing mass at 0.37~kg/s, corresponding to a production rate of around $8\times10^{24}$~N$_2$ molecules per second and a rate of change in the radius of 0.9~cm/day.  For comparison \citet{trilling2018} place $3\sigma$ upper limits on the production of dust ($< $9~kg/s), CO$_2$ ($< 9\times10^{22}$~molecules per second), and CO ($< 9\times10^{21}$~molecules per second).  The mass loss rate we calculate is well below the dust production limit of \citet{trilling2018} so if dust makes up some fraction of the material ejected it would not be detectable.  The constraints on CO$_2$ and CO, on the other hand, are much more restrictive, at only 1\% and 0.1\% of the N$_2$ production rate respectively, indicating that CO and CO$_2$ cannot represent more than minor trace impurities in the N$_2$ ice.  We are not aware of any constraints that have been placed on CH$_4$, and N$_2$ does not have strong spectral lines in the infrared, making it hard to detect.  N$_2$ outgassing would likely be best observed in the ultraviolet.

\citet{demeo2010} and \citet{merlin2010} constrain the fraction of CO dissolved in N$_2$ ice on Pluto to be no more than around 0.1\%.  If `Oumuamua had a similar fraction of CO dissolved the loss rate could be up to $\sim8\times10^{21}$~CO molecules per second, which is compatible with the constraints from \citet{trilling2018}.  As we noted above, CH$_4$ is observed on the surface of Pluto as a trace species dissolved in N$_2$ ice at levels no more than a few weight percent \citep{trafton2015} and we feel some trace impurities of CH$_4$ are likely to explain the red colour of `Oumuamua.  `Oumuamua is as red as some of the reddest Solar System objects, with optical spectral slopes measured variously as $10\pm6$\%/100~nm \citep{ye2017}, $22\pm15$\%/100~nm \citep{bannister2017}, or $17\pm2.3$\%/100~nm ($9.3\pm0.6$\%/100~nm) \citep[][ACAM (X-shooter)]{fitzsimmons2018}.  With the exception of the \citet{fitzsimmons2018} X-shooter data these are consistent with the spectral slopes of Pluto \citep{lorenzi2016}, Makemake \citep{licandro2006} and Gonggong \citep{santos-sanz2012}, all around 14\%/100~nm and presumably attributable to photolyzation of CH$_4$ to form tholins.

In terms of other possible trace species, \citet{ye2017} used ground-based observations with the Hale Palomar telescope to place upper limits on the production rates of CN ($< 2\times10^{22}$~molecules per second) and C$_2$ ($< 4\times10^{22}$~molecules per second), but these are not expected in the N$_2$ ices that make up the crust of a Pluto-like body.  Radio observations by \citet{park2018} with the Green Bank Telescope place a loose upper limit on the production rate of OH ($< 1.7\times10^{27}$~molecules per second), but this is much higher than our predicted production rate of N$_2$.  Far more restrictive for OH production is the extremely poor acceleration provided by sublimation of H$_2$O such that compared to N$_2$ it effectively acts as an inert diluent.

\section{Temporal evolution of `Oumuamua}
\label{sec:temporal}

We demonstrated above that a pure N$_2$ ice composition is capable of reproducing the observed non-gravitational acceleration at a distance of around 1.4~au from the Sun for an albedo of 0.64.  At that time, however, `Oumuamua would have had a mass of only $8.01 \times 10^6$~kg while losing mass at a rate of 0.37~kg/s ($3.2 \times 10^4$~kg/day) such that even between 1.4 and 2~au it would have shrunk by around 0.7~m.  It is thus clear that `Oumuamua would have evolved substantially over time.

\subsection{Passage through the solar system}
\label{sec:temporal:solar}

\begin{figure}
    \centering
    \includegraphics[width=\columnwidth]{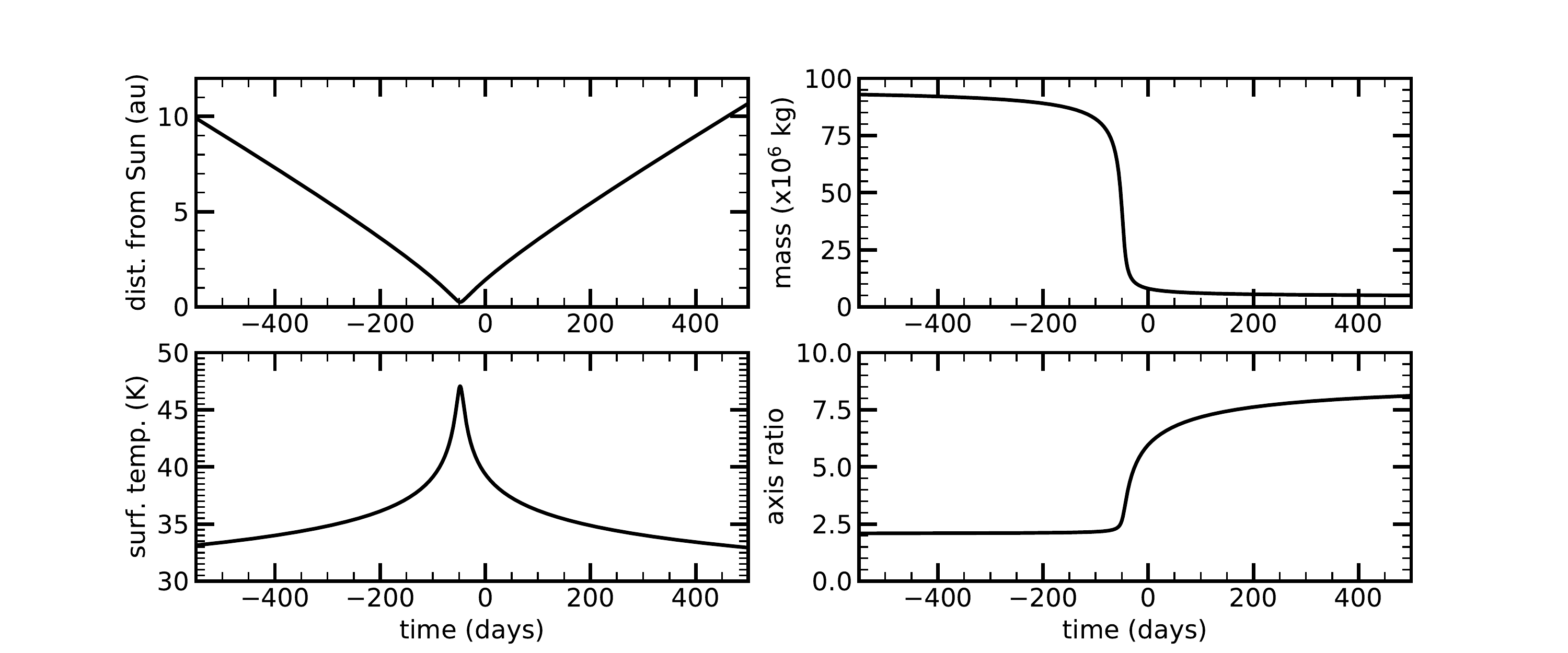}
    \caption{Change in `Oumuamua's parameters over time.  Clockwise from top left: distance from the Sun, mass, axis ratio ($a/c$), surface temperature.  Times are relative to 27 October 2017.  Perihelion occurs at -48 days.}
    \label{fig:timeplot}
\end{figure}

As we can see from Equation~\ref{eq:dmdt}, the rate of mass loss from `Oumuamua is a strong function of the heliocentric distance.  From our starting point at 1.42~au we can integrate forwards and backwards to find the evolution of the size and shape of `Oumuamua over time.  In Figure~\ref{fig:timeplot} we show the evolution in the mass, surface temperature, and axis ratio ($a/c$) of `Oumuamua over a period of around 18 months before and after perihelion alongside its distance from the Sun for comparison.  Times are measured relative to 27 October 2017 at which epoch we fix the distance from the Sun as 1.42~au and the size as 45.5~m $\times$ 43.9~m $\times$ 7.5~m.  The albedo is set at 0.64, which we assume remains constant.  For the orbital evolution we assume a semi-major axis of -1.2978~au and eccentricity 1.19951 for which perihelion passage occurs 48 days before our fixed point (9 September 2017)\footnote{From JPL HORIZONS service https://ssd.jpl.nasa.gov/sbdb.cgi?sstr=2017U1}.  The non-gravitational acceleration is never large enough to modify the orbit sufficiently to significantly alter our calculations, and so for the purposes of Figure~\ref{fig:timeplot} we neglect the non-gravitational acceleration.  We assume that in a time interval $\Delta t$ each semi-axis of the triaxial ellipsoid decreases by an amount $h=(dR/dt)\Delta t$, and the surface area of the ellipsoid is re-computed at the end of each timestep.

Unsurprisingly, the period immediately around perihelion dominates the change in mass and axis ratio.  In the 50 days either side of perihelion passage the mass drops by a factor of 10 while the axis ratio rises from just above 2:1 to 6:1.  It is notable, however, that the evaporative cooling due to the extreme mass loss is so effective that the surface never rises above 47~K.  Outside this narrow window, mass loss continues at a much lower rate, and the axis ratio continues to undergo significant evolution because the $c$ axis has shrunk to such a small size.  By the time `Oumuamua passed the orbit of Uranus in September 2020 it would have dropped in mass by roughly another factor of 2, to $4.81 \times 10^6$~kg, and reached an axis ratio of 8.3:1.

\begin{figure}
    \centering
    \includegraphics[width=\columnwidth]{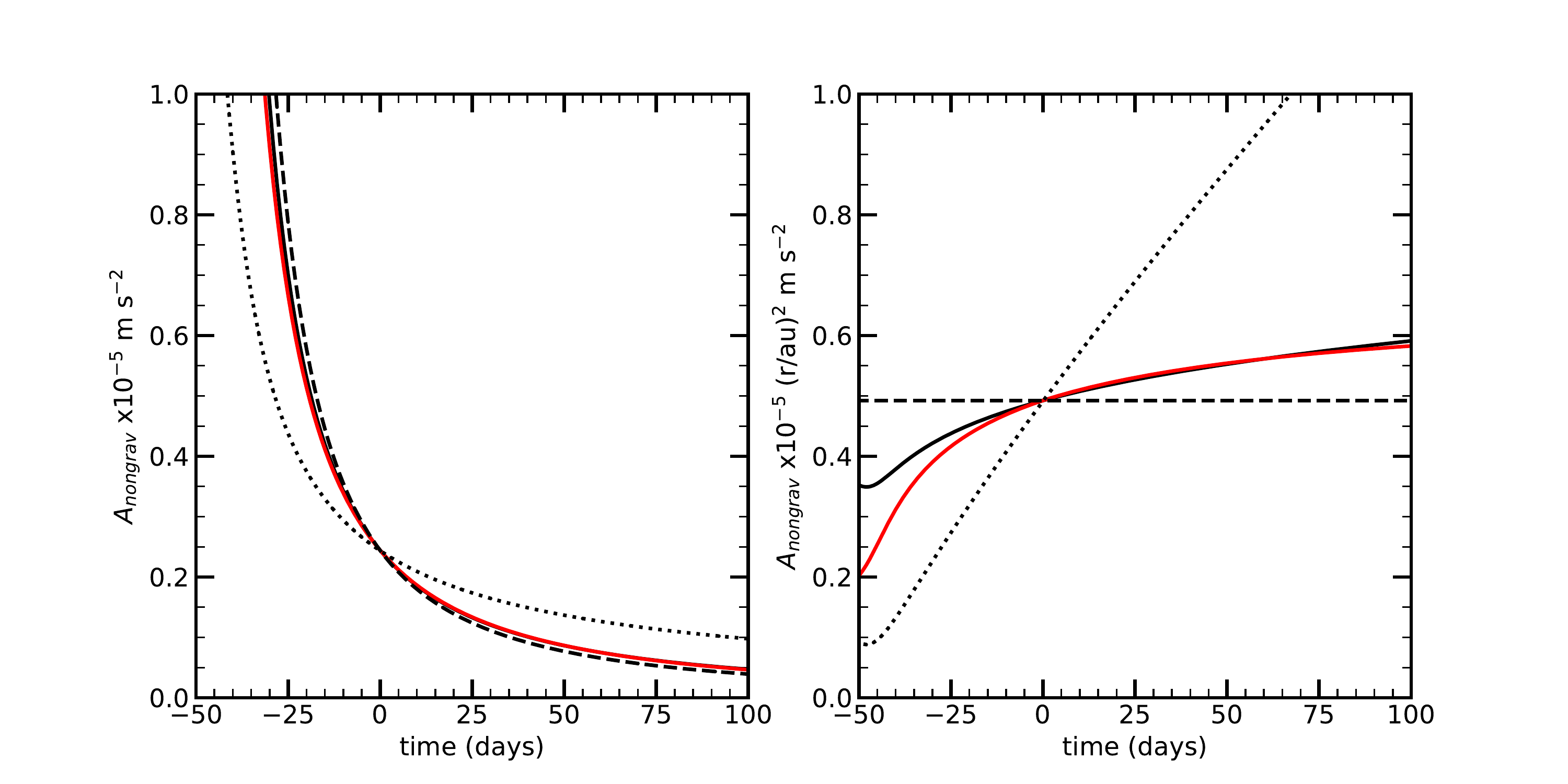}
    \caption{Non-gravitational acceleration predicted by our model (solid red) for time around 27 October 2017 as compared with different relationships that are constant power laws in $d$.  At left we show the predicted non-gravitational acceleration while at right we multiply by ($d$/au)$^2$ to provide a more detailed view of the differences between the curves.  The dashed and dotted black lines show $d^{-2}$ and $d^{-1}$ relations respectively, while the solid black line shows a $d^{-1.8}$ relationship, which provides a good fit to our prediction over the relevant range.  The range of observations over which the acceleration was fit by \citet{micheli2018} runs from -13 to +68 days (14 October 2017 to 2 January 2018).}
    \label{fig:nongravtime}
\end{figure}

As the mass and mass loss changed, so too would have the non-gravitational acceleration.  \citet{micheli2018} found that the observed non-gravitational acceleration obeyed a relationship that lies between $d^{-1}$ and $d^{-2}$, probably closer to $d^{-2}$.  We can immediately see from Equation~\ref{eq:nongravacc} that the insolation induces a $d^{-2}$ dependence, with a non-linear deviation for the thermal emission term.  In Figure~\ref{fig:nongravtime} we show the change in our predicted non-gravitational acceleration as a function of time during a window of a few months spanning the range of observations that were used by \citet{micheli2018} to fit the non-gravitational acceleration, these observations were densest from 18 October to 23 November 2017 with precovery data for 14 October 2017, and additional points on 12 December 2017 and 2 January 2018.  Not only does our predicted acceleration match the magnitude of the non-gravitational acceleration at 1.42 au, but over the relevant range our predicted curve is very close to $d^{-1.8}$, which we consider a good fit to the observations given the uncertainties involved, both in the observations and in some of the parameters in our equations (e.g., what value of $\tau$ is appropriate, and corrections for the ellipsoidal shape of `Oumuamua).

Our scenario is also consistent with `Oumuamua’s rotation. `Oumuamua is tumbling, in non-principal axis rotation \citet{fraser2018}, but spinning only once per 8 hours or so. This `slow' rotation rate of `Oumuamua has been taken by \citet{rafikov2018b} as evidence against outgassing, arguing that the forces needed to provide the non-gravitational acceleration would torque and spin up the object until it underwent rotational fission. The underlying picture is one like a normal comet, in which most of the jetting occurs from isolated spots on the surface, providing a torque that increases the rotation rate $\Omega$ at a rate $d\Omega/dt = \zeta (R M /I) a$, where $a$ is the non-gravitational acceleration, $R$, $M$ and $I$ the characteristic size, the mass, and the moment of inertia of the object, $\zeta$ a dimensionless number such that $\zeta R$ is the effective lever arm.

Based on his analysis of 7 regular comets, \citet{rafikov2018a} derived an average value of $\log \zeta = -2.21 \pm 0.54$. \citet{mashchenko2019} found a best fit of `Oumuamua’s light curve if it were oblate and experienced a torque consistent with $\log \zeta = -2.34$, which, as they pointed out, is within the range of values \citet{rafikov2018a} inferred for comets. Since $I \sim M R^2$, the spin-up rate is proportional to $R^{-1}$ and therefore $p^{1/2}$. For a given spin-up rate, $\zeta$ is proportional to $p^{-1/2}$. Because we are arguing for an albedo $p = 0.64$ rather than the value $p = 0.1$ assumed by \citet{mashchenko2019} when deriving $\zeta$, we favour a smaller body more easily spun up, and therefore a smaller value of the effective lever arm coefficient, $\log \zeta \approx -2.74$. However, this is still within the range of values observed among comets \citet{rafikov2018a}. Moreover, a lower value of $\zeta$ is consistent with the idea of `Oumuamua as a monolith of pure N$_2$ ice without localised jetting, in accordance with our assumption of sublimation across the hemisphere, and our use of the coefficient 1/3, in our derivation of the jetting force.  We note that \citet{seligman2019} also examined the possible spin-up of `Oumuamua and found that it is possible for the rotation rate to oscillate around the observed $\sim$8 hour value under the action of sublimation jetting for appropriate venting angles, however it is not clear how their results scale to a body with a substantially different albedo. It has been argued that for `Oumuamua to aquire its tumbling motion would require many Gyr of passage through the interstellar medium \citep{zhou2020}, but it seems clear that `Oumuamua must have experienced significant (albeit lower than is typical for comets) torques within the Solar System, as it lost $\sim$92\% of its mass.

\subsection{Passage through the interstellar medium}
\label{sec:temporal:ism}

An important factor in ruling out H$_2$ as a likely composition for `Oumuamua is that an H$_2$ ice body would experience rapid erosion during its passage through the interstellar medium (ISM).  \citet{hoang2020} discussed this erosion in detail, showing that even a multi-km H$_2$ ice body would be completely eroded away in less than $10^8$~years.  For H$_2$, \citet{hoang2020} found that simple thermal sublimation was dominant since the equilibrium surface temperature of a body in the ISM is barely below the sublimation temperature of H$_2$ ice.  The sublimation temperature of N$_2$ ice is about a factor of 7 higher than that of H$_2$ ice and the exponential dependence of sublimation (Eq.~\ref{eq:drdt}) makes it immediately apparent that direct, thermally driven sublimation will be negligible at ISM temperatures for N$_2$; but it is nonetheless prudent to consider other possible erosion mechanisms.

\citet{domokos2009} described how isotropic abrasion by dust grains impacting the surface of a body can lead to an increase in the axis ratio, much as we have described above for outgassing, and \citet{domokos2017} attributed `Oumuamua's shape to abrasion by dust grains eroding its surface as it passed through the ISM. However, `Oumuamua is unlikely to encounter sufficient material as it passes through the ISM to result in significant change to its mass and dimensions.  For a typical ISM density of around 1 proton per cm$^3$ and a dust-to-gas mass ratio of 0.01, a body with mean diameter $\sim$50~m and a relative velocity $\sim$10~km/s will only collide with around $10^3$~kg per Gyr of matter in total, and only around 10~kg per Gyr of dust, a tiny fraction of the $\sim10^7$~kg mass of the body.  Even travelling 10~pc through a giant molecular cloud with a mean density of $10^3$ protons per cm$^3$ would only result in encountering around $10^3$~kg of dust.  Dust abrasion is thus clearly insufficient to result in any change to `Oumuamua's size or shape.

Another possible mechanism is photodesorption.  While both visible and UV photons have sufficient energy to overcome the desorption energy of an N$_2$ molecule in N$_2$ ice ($\sim$0.07 eV) the efficiency of the process is only around $5\times10^{-3}$ N$_2$ molecules per photon, since N$_2$ ice is an inefficient absorber at most optical and UV wavelengths \citep{bertin2013, fayolle2013}.  We assume the interstellar radiation field will deliver photons with an energy flux of $2.7\times10^{-6}$~W~m$^{-2}$, each with a typical energy of around 10~eV \citep{mathis1983}.  This would produce a desorption rate of $8\times10^9$~N$_2$ molecules~m$^{-2}$~s$^{-1}$, or an erosion rate of just over 1~cm/Gyr.  As with dust abrasion, this is insufficient to result in any substantial changes in `Oumuamua's size.

Finally we consider galactic cosmic rays (GCRs).  Integrating over energy, the GCR proton and alpha-particle energy flux in the ISM near the Sun is around $1.9\times10^{-5}$~W~m$^{-2}$ [using the analytical formula of \citet{webber1998}, but scaled down by a factor of 1.5 to better match the observations compiled by \citet{tatischeff2014}].  The majority of the incident particles would have energies in the range 10-100 MeV/nucleon.  \citet{vasconcelos2017} measured the erosion of an N$_2$:CH$_4$ ice (95:5 mass ratio) by 15.7 MeV oxygen ions (1 MeV/nucleon) and found that after receiving a fluence of $6\times10^{17}$~ions~m$^{-2}$ the ice was reduced in thickness by about 8~$\mu$m.  The ions deposited 930~keV/$\mu$m, or a total of 7.4~MeV each.  Based on this experimental data we infer the removal of 1 N$_2$ molecule for roughly every 26~eV delivered by GCRs.  We note that the stopping lengths of typical GCRs will not exceed a fraction of a cm and so assume there is no reduction in erosion efficiency for higher energy GCRs.  For an interstellar GCR energy flux of $1.9\times10^{-5}$~W~m$^{-2}$ we calculate an average erosion rate of 6.6~m/Gyr.  For N$_2$ ice, GCR erosion is thus the dominant erosion mechanism in the ISM and can potential alter the size and shape of the body over long periods.  This erosion rate due to GCRs is about the same as we calculate for the thermal sublimation due to solar radiation at a distance of about 130~au, so beyond around 130~au from the Sun GCR erosion dominates over thermal sublimation.  This is roughly the same distance as the heliopause, within which GCRs are suppressed by the Solar magnetic field \citep{gurnett2013}.

It is important to note that the GCR erosion rate of 6.6~m/Gyr corresponds to the GCR flux in the neighbourhood of the Sun, today.  The GCR flux is roughly proportional to the star formation rate, and so the GCR flux can be expected to have tracked variations in the star formation rate in the vicinity of the Sun over time.  The Sun is currently in an inter-arm region with a GCR flux characteristic of much of the Galaxy, but the star formation rate in the spiral arms is substantially higher, such that the GCR flux is expected to be around 3.7 times higher within a spiral arm than the galactic average \citep{dunham2020, fujimoto2020}.  Over a period of a Gyr, travelling at 9~km/s relative to the local standard of rest, `Oumuamua could travel tens of kpc, likely passing in and out of spiral arms multiple times.  The analysis of \citet{vallee2005} suggests that the Sun spends roughly 50\% of its time within 1 kpc of spiral arms (the diffusion length of GCRs), and the other 50\% in the inter-arm regions.  This implies an average erosion rate over the last few hundred Myr that is around a factor of 2.4 times higher than the rate in the Solar neighbourhood today, i.e. $\sim$15.4~m/Gyr.  In addition, the star formation rate was higher in the past across the entire Galaxy: it was greater than 5 times the present rate over 8~Gyr ago, falling to about twice the current rate 5-6 Gyr ago, peaking again at around 5 times the present value 2-3 Gyr ago, and finally falling since then to the present value \citep{mor2019}.  Over the 4.5 Gyr since the birth of the Solar system the total erosion would have been 260~m along each semi-axis, averaging 57~m/Gyr.  In the last 2~Gyr `Oumuamua would have eroded by 92~m at an average of around 3 times the present rate, and even over just the last Gyr it would have eroded at an average of about twice the present rate (including the correction for spiral arms) for a total of around 31~m.

\begin{table}
    \centering
    \caption{Physical properties of  `Oumuamua at selected epochs.}
    \begin{tabular}{p{2 cm}|p{1.5 cm}|c|c|c|c|c|c|l}
        Epoch & Distance from Sun & Time & $T_{\rm surf}$ & 2$a$ & 2$b$ & 2$c$ & $a/c$ & mass \\
         & (au) & & (K) & (m) & (m) & (m) & & ($\times10^6$~kg) \\
        \hline
        Ejection from parent system & - & $\sim$0.45~Gyr ago & - & 92.4 & 90.8 & 54.4 & 1.70 & 244 \\
         & 130  & 14 Apr. 1995 & 19.9 & 72.4 & 70.8 & 34.5 & 2.10 & 94.4 \\
         & 30   &  1 Dec. 2012 & 30.1 & 72.4 & 70.8 & 34.4 & 2.10 & 94.2 \\
         & 5.2  & 16 Jan. 2017 & 35.0 & 71.7 & 70.1 & 33.8 & 2.12 & 90.6 \\
        Perihelion passage         & 0.255&  9 Sep. 2017 & 47.1 & 57.6 & 56.1 & 19.7 & 2.92 & 34.0 \\
        Optical observations       & 1.42 & 27 Oct. 2017 & 39.4 & 45.4 & 43.9 &  7.50 & 6.06 & 7.98 \\
        \emph{Spitzer} observations& 2.0  & 21 Nov. 2017& 38.1 & 44.7 & 43.2 &  6.81 & 6.57 & 7.03 \\
         & 5.2  &  2 May 2018  & 35.0 & 43.6 & 42.0 &  5.64 & 7.72 & 5.52 \\
         & 30   & 18 June 2022 & 30.1 & 42.9 & 41.3 &  4.96 & 8.66 & 4.65 \\
         & 130  & 14 Feb. 2040 & 19.7 & 42.8 & 41.3 &  4.92 & 8.72 & 4.64 \\
    \end{tabular}
    
    \label{tab:evol}
\end{table}

How long `Oumuamua has been travelling through the ISM is not known, but \citet{almeidafernandes2018} placed an upper limit of around 1.9 - 2.1~Gyr based on its low velocity dispersion relative to the LSR.  Had `Oumuamua been in interstellar space for this maximum time of around 2~Gyr, it would have been eroded by over 90~m in radius, implying an initial mass upon ejection from its stellar system of around $8 \times 10^{9}$~kg.  This would mean that `Oumuamua entered the solar system with only 1\% of the mass it had when it left its parent system; while this is not impossible, it seems unlikely.  By comparison, if `Oumuamua departed its parent system around 0.4-0.5~Gyr ago, it would have been eroded by around 10~m along each semi-axis, entering the Solar system with slightly under half of its initial mass, a much more plausible value.  Travelling at 9~km/s for around 0.4-0.5~Gyr, `Oumuamua could have travelled about 4~kpc, albeit not in a straight line: its motion through the Galactic potential would have changed its velocity en route, and this distance must include epicyclic motions.  Since a young stellar system is the most likely candidate to be ejecting large quantities of material we tentatively suggest an origin around 0.4-0.5~Gyr ago in the Perseus spiral arm, which is about 2-3~kpc from the Sun \citep{kounkel2020} and consistent with `Oumuamua's approach from the direction of Vega.  Using this starting point we provide a summary of the mass and dimensions of `Oumuamua at various epochs along its journey from its parent system, to, and through the Solar system in Table~\ref{tab:evol}.
We note that if `Oumuamua were travelling through the ISM for 0.4-0.5 Gyr, it would have seen its axis ratios increase from $c/a=1.7$ to 2.1. Typical axis ratios of fragments in the solar system are $a:b:c \sim 2:1.4:1$, never exceeding 3:1 \citep{domokos2017}. While it would not be implausible for `Oumuamua to have been ejected from its stellar system with an axis ratio of 2.1, an initial axis ratio 1.7:1, consistent with travel through the ISM for 0.4-0.5 Gyr, is apparently more likely.

\section{Summary}
\label{sec:summary}

We have proposed that 1I/`Oumuamua was a fragment of nearly pure N$_2$ ice.  We have shown that this is consistent with the observation contraints on `Oumuamua's size, albedo ($p$), composition, and non-gravitational acceleration.  We highlight that the acceleration scales as the product of the sublimation rate, proportional to $(1-p)$, and the surface area per unit mass, proportional to $p^{1/2}$.  Therefore there are generally two values of the albedo that may satisfy the equations: one with low albedo, $p\sim0.1$, and one with high albedo, $p\sim0.7$.  The former is consistent with most small bodies in the Solar system, but the latter is consistent with the surfaces of Pluto and Triton, which importantly have surfaces dominated by N$_2$ ice.

The non-gravitational acceleration of `Oumuamua is exactly consistent with that inferred by \citet{micheli2018} if `Oumuamua is a solid chunk of pure N$_2$ ice, with albedo 0.64 and axes (at the time of observation) 45.4 $\times$ 43.9 $\times$ 7.50~m.  This shape is consistent with the oblate spheroid solution of \citet{mashchenko2019}, re-scaled to our albedo.  Nearly pure N$_2$ ice is observed in abundance on the surfaces of Pluto and Triton, with the surface of Pluto being 98\% N$_2$ ice.  This composition would be consistent with the lack of dust emission, and the small amounts of CO dissolved in N$_2$ ice found on Pluto ($\sim$0.1~wt\%) would not violate the constraints on CO production set by the \emph{Spitzer} observations of \citet{trilling2018}.  The trace amounts of CH$_4$ typically found in N$_2$ ice on Pluto (a few wt\%) would photolyze to produce tholins, reddening the surface of `Oumuamua in the same way as the surface of Pluto; both bodies are consistent with a spectral slope of around 14\%/100~nm.  A fragment of N$_2$ ice matching the N$_2$ ice found on the surface of Pluto thus exactly matches all of the observational constraints on `Oumuamua, and may be the only known material found in the Solar system that can do so.

Our modelling shows that, perhaps surprisingly, an N$_2$ ice fragment can survive passing the Sun at a perihelion distance of 0.255~au, in part because evaporative cooling maintains surface temperatures less than 50~K.  Despite being closer to the Sun than Mercury, `Oumuamua's surface temperatures remained closer to those of Pluto.  The volatility of N$_2$ did, however, lead to significant mass loss - we calculate that by the time `Oumuamua was observed, a month after perihelion, it retained only around 8\% of the mass it had on entering the solar system.  This loss of mass is key to explaining the extreme shape of `Oumuamua: isotropic irradiation and removal of ice by sublimation increases the axis ratios, a process also identified by \citet{seligman2020}.  Between entering the Solar system and the light curve observations the loss of mass from `Oumuamua increased its axis ratios from an unremarkable 2:1 to the extreme observed value of around 6:1.

Many more exotic explanations have been proposed to explain various of the properties of `Oumuamua.  Proposals to explain the non-gravitational acceleration have included extremely low density fractal aggregates of water ice \citep{flekkoy2019, moromartin2019, luu2020}, H$_2$ ice \citep{seligman2020}, and solar sails \citep{bialy2018}.  None of these have ever been observed.  The extreme axis ratios have also been the subject of a large number of proposed explanations.  For example, \citet{katz2018} suggested that `Oumuamua was ejected from a system in which the host star was entering the red giant stage, and that the enhanced luminosity of the star heated the body and fluidized it, enabling it to achieve a prolate Jacobi ellipsoid shape.  However, this would demand a high density incompatible with the presence of volatiles and the peculiar velocity of an old star entering the red giant stage would be inconsistent with the observed low velocity of `Oumuamua relative to the local standard of rest.  Meanwhile, \citet{cuk2018} proposed that the extreme shape might be consistent with tidal effects as a massive body passed close to a star and was disrupted, but it is not clear that such an event can reproduce the axis ratios derived by \citet{mashchenko2019}, and in any case tidal disruptions like this would be rare \citet{jackson2018}.

A key advantage of the proposal we advance here of an N$_2$ ice fragment is that it can simultaneously explain all of the important observational characteristics of `Oumuamua, and that material of this composition is found in the solar system.
We therefore conclude that 'Oumuamua is an example of an uncommon but certainly not exotic object: a fragment of a differentiated Pluto-like planet from another stellar system.
In the companion paper (Desch and Jackson, 2021) we examine whether N$_2$ ice fragments the size of `Oumuamua would be ejected from the surfaces of ``exo-Plutos" with sufficient frequency to explain this unusual object.


%
%
%
%
%
%
%
%

\acknowledgments
We thank Yusuke Fujimoto, Chris Glein, Greg Laughlin, and Darryl Seligman for useful discussions.  We thank Sean Raymond and the anonymous reviewer for useful comments that helped us to improve the clarity of the manuscript.  The results reported herein benefitted from collaborations and/or information exchange within NASA's Nexus for Exoplanet System Science (NExSS) research coordination network sponsored by NASA's Science Mission Directorate.

The computer code used to generate all figures and data in this manuscript are available at \citet{jackson2021}\footnote{ \url{https://doi.org/10.5281/zenodo.4558478}}.

\clearpage
\appendix
\section{Ice thermal data}
\label{sec:appendix}

\begin{sidewaystable}
  \caption{Data for hydrogen, neon and astrophysical ices.  From left to right the data columns are sublimation temperature, Tsub; density, $\rho$; sublimation enthalpy, $\Delta {\rm H_{sub}}$; enthalpy for any solid state phase transition that occurs in the relevant temperature range, $\Delta {\rm H_{trans}}$; heat capacity, C$_{\rm p}$; thermal conductivity, $k$; lattice vibrational frequency, $\nu$.  Each data column is followed by a reference column listing the sources for the data.  The Nitrogen $\alpha-\beta$ phase transition occurs at 35.6 K, while oxygen has two transitions in the relevant range, $\alpha-\beta$ at 23.9 K and $\beta-\gamma$ at 43.8 K.  We list errors (where available) only for the sublimation enthalpy since this is by far the dominant contribution to errors on the curves in Figure 1 where multiple references are listed we take the mean and list the standard deviation of the values as the error.}
   \label{tab:icedata}
   \begin{tabular}{c|c|c|c|c|c|c|c|c|c|c|c|c|c|c}
    Ice   & $T_{\rm sub}$ & ref &   $\rho$  & ref & $\Delta {\rm H_{sub}}$ & ref & $\Delta {\rm H_{trans}}$ & ref & C$_{\rm p}$ & ref & $k$ & ref & $\nu$ & ref \\
        &       K       &     &kg m$^{-3}$&     &       kJ/mol           & &      J/mol               &     &   kJ/K/kg   &     &W/m/K&     & s$^{-1}$ & \\
    \hline
    H$_2$ &  4  &  1  &   86 &  1  & 0.85          & 2                 &  -  &  -  & 0.8   & 2  & 1   & *2,3  & 7.5 &  4 \\
    Ne    &  9  &  5  & 1444 &  6  & 1.9$\pm$0.29  & 5                 &  -   &  - & 0.207 & 7  & 0.3 & 8     & 1.4 &  9 \\
    N$_2$ & 25  &  5  & 1020 &  5  & 6.85$\pm$0.36 &$^+$5,10,11,12,13,14& 215 & 15 & 1.225 & 16 & 0.3 & 17,18 & 6.5 & 14 \\
    CO    & 29  & 19  &  930 & 20  & 7.3$\pm$0.6   & 19                &  -   &  - & 7.64  & 21 & 0.7 &    22 & 2   & 23 \\
    O$_2$ & 31  &  5  & 1530 &  5  & 9.26$\pm$0.42 & 5                 & 90 ($\alpha-\beta$) & 24,25 & 0.87 & 25 & 0.35 & 25 & 1.5 & 26 \\
          &     &     &      &     &               &                   & 750 ($\beta-\gamma$) & & & & & & & \\
    CH$_4$& 36  & 19  &  520 & 27  & 9.4$\pm$0.7   & 19                &  -   &  - & 1.2   & 28 & 0.3  & 29 & 3.5 & 30 \\
    CO$_2$& 82  &5,19 & 1560 &  5  & 26.5$\pm$2.33 & 5,19,31           &  -   &  - & 0.83  & 31 & 0.2  & 17 & 2.9 & 32 \\
    NH$_3$& 100 & 19  &  817 & 33  & 28.8          & 19                &  -   &  - & 1.38  & 34 & 4    & 35 & 3.5 & 36 \\
    H$_2$O& 158 &  5  &  920 &  -  & 49.58$\pm$5.24& 5,23              &  -   &  - & 1.27  &37,38& 4.3 & 39 & 2   & 23 \\
   \end{tabular}
   \begin{tabular}{p{7.5cm} p{7.5cm} p{7.5cm}}
    *The thermal conductivity of solid hydrogen
    & 13) \citet{collings2015}
    & 27) \citet{ramsey1963}\\
    
    is strongly dependent on the ratio of
    & 14) \citet{fayolle2016},
    & 28) \citet{vogt1976}, at 21~K \\
    
    ortho-hydrogen to para-hydrogen
    & for $^{15}$N$_2$, $\nu$ scaled by $m^{-1/2}$
    & 29) \citet{jezowski1997}, at 20~K \\
    
    $^+$We exclude the value found by 
    & 15) \citet{lipinski2007}
    & 30) \citet{orbriot1978}, from peak in Raman \\
    
    \citet{luna2014} as an outlier
    & 16) \citet{trowbridge2016}, at 30~K
    & spectrum at 118~cm$^{-1}$ \\
    
    1) \citet{silvera1980}
    & 17) \citet{cook1976}, at 30~K
    & 31) \citet{giauque1937}, heat capacity at 82~K \\
    
    2) \citet{souers1986}, at 4.2~K
    & 18) \citet{stachowiak1994}, at 30~K
    & \\
    
    3) \citet{huebler1978}, at 6~K
    & 19) \citet{luna2014}, and references therein
    & 32) \citet{sandford1990} \\
    
    4) \citet{sandford1993b}
    & 20) \citet{bierhals2001}
    & 33) \citet{blum1975} \\
    
    5) \citet{shakeel2018}, and references therein
    & 21) \citet{clayton1932}
    & 34) \citet{overstreet1937} \\
    
    6) \citet{hwang2005}
    & 22) \citet{stachowiak1998}
    & 35) \citet{romanova2013} \\
    
    7) \citet{fenichel1966}, at 9~K
    & 23) \citet{sandford1988}
    & 36) \citet{sandford1993a} \\
    
    8) \citet{weston1984}, at 9~K
    & 24) \citet{Szmyrka1998}
    & 37) \citet{giauque1936} \\
    
    9) \citet{gupta1969}
    & 25) \citet{freiman2004}, and references
    & 38) \citet{shulman2004}\\
    
    10) \citet{frels1974}
    & therein, heat capacity, thermal conductivity
    & 39) \citet{slack1980} \\
    
    11) \citet{oberg2005}
    & at 30~K, $\beta$ phase
    & \\
    
    12) \citet{bisschop2006}
    & 26) \citet{bier1984}, from peak in Raman
    & \\

    & spectrum at 50cm$^{-1}$
    & \\
    
   \end{tabular}
\end{sidewaystable}


%
%

\bibliography{refs}

%
%
%
%
%

\end{document}